\documentclass[aps,reprint]{revtex4-2}

\usepackage{amsmath,amssymb}
\usepackage{graphicx,graphics}
\usepackage{dcolumn}
\usepackage{bm}
\usepackage{color}
\usepackage{lipsum}

\begin{document}

\title{Tunneling magnetoresistance and spin-valley polarization of aperiodic magnetic silicene superlattices}

\author{P. Villasana-Mercado}
\affiliation{Unidad Académica de Ciencia y Tecnología de la Luz y la Materia, Universidad Autónoma de Zacatecas,
Carretera Zacatecas-Guadalajara Km. 6, Ejido La Escondida, 98160 Zacatecas, Zacatecas, Mexico.}%

\author{J. G. Rojas-Briseño}
\affiliation{Unidad Académica de Ciencia y Tecnología de la Luz y la Materia, Universidad Autónoma de Zacatecas,
Carretera Zacatecas-Guadalajara Km. 6, Ejido La Escondida, 98160 Zacatecas, Zacatecas, Mexico.}%

\author{S. Molina-Valdovinos}
\affiliation{Unidad Académica de Ciencia y Tecnología de la Luz y la Materia, Universidad Autónoma de Zacatecas,
Carretera Zacatecas-Guadalajara Km. 6, Ejido La Escondida, 98160 Zacatecas, Zacatecas, Mexico.}%

\author{I. Rodríguez-Vargas}
\email{Corresponding author email: isaac@uaz.edu.mx}
\affiliation{Unidad Académica de Ciencia y Tecnología de la Luz y la Materia, Universidad Autónoma de Zacatecas,
Carretera Zacatecas-Guadalajara Km. 6, Ejido La Escondida, 98160 Zacatecas, Zacatecas, Mexico.}%

\date{\today}

\begin{abstract}
Magnetic silicene superlattices (MSSLs) are versatile structures with spin-valley polarization and tunneling magnetoresistance (TMR) capabilities. However, the oscillating transport properties related to the superlattice periodicity impede stable spin-valley polarization states reachable by reversing the magnetization direction. Here, we show that aperiodicity can be used to improve the spin-valley polarization and TMR by reducing the characteristic conductance oscillations of periodic MSSLs (P-MSSLs).  Using the Landauer-Büttiker formalism and the transfer matrix method, we investigate the spin-valley polarization and the TMR of Fibonacci (F-) and Thue-Morse (TM-) MSSLs as typical aperiodic superlattices.  Our findings indicate that aperiodic superlattices with higher disorder provide better spin-valley polarization and TMR values.  In particular,  TM-MSSLs reduce considerably the conductance oscillations giving rise to two well-defined spin-valley polarization states and a better TMR than F- and P-MSSLs. F-MSSLs also improve the spin-valley polarization and TMR, however they depend strongly on the parity of the superlattice generation. 
\end{abstract}


\maketitle


\section{Introduction}
Currently, it is required to have electronic devices with a large storage and information processing capacity that allow us to cover our daily needs~\cite{shen2008spintronics}. Then, it would be useful to have versatile devices, which could be used for processing and reading information. To process information, it is required to have two fully defined polarization states to make binary logic. This can be achieved by means of spintronics and valleytronics by manipulating the spin and valley electron degrees of freedom, respectively~\cite{sierra2021van,hirohata2020review,schaibley2016valleytronics}. On the other hand, the reading of information on hard drives was developed thanks to the discovery of giant magnetoresistance~\cite{PhysRevB.39.4828,PhysRevLett.61.2472} and tunneling magnetoresistance (TMR)~\cite{JULLIERE1975225,PhysRevLett.74.3273}. TMR occurs when ferromagnetic layers are intercalated with non-magnetic and non-metallic layers, where one of the ferromagnetic layers can vary its magnetization. 
A change in resistance takes place when the magnetization configuration changes from parallel (PM) to antiparallel (AM), that is, when the ferromagnetic layers have the same magnetization direction or opposite. 
 This is known as a spin valve and is used to sense the individual magnetization domains, which correspond to bits, on a hard disk drive. The arrival of two-dimensional materials with novel and exotic properties that are different from their bulk counterparts makes them an ideal candidate for electronic applications. Specially, the spin and valley degrees of freedom, and the intrinsic magnetic properties of some 2D materials are favorable for the creation of versatile devices~\cite{doi:10.1063/5.0032538,ahn2020,RevModPhys.92.021003,2Dmaterials}. 

In this context, silicene is ideal for spintronics and valleytronics due to its exotic properties~\cite{chen2012evidence,PhysRevLett.107.076802,ezawa2012valley} and its compatibility with silicon-based nanotechnology. Silicene is made by silicon atoms alternate in two parallel planes. This particular configuration of the atoms promotes the  $sp^{2}-sp^{3}$ hybridization giving rise to the silicene's buckled geometry~\cite{PhysRevB.88.035432}. Silicene was first proposed theoretically~\cite{PhysRevB.50.14916,PhysRevLett.102.236804} and later synthesized in the laboratory~\cite{doi:10.1063/1.3524215,PhysRevLett.108.155501}. In addition, this material has a band gap at Dirac (K, K$'$) points that can be modulated by an applied external electric field~\cite{Liu_2014,zni2012,Ezawa_2012}. Pristine silicene exhibits a large spin–orbit coupling (SOC) suitable for spin–valley quantum effects~\cite{ZHAO201624}. The spin inversion symmetry between the valleys K and K$'$ induces spin-valley currents~\cite{Shakouri2015}. Several strategies have been implemented to manipulate the spin and valley polarized currents~\cite{Yokoyama2013,Farokhnezhad2017,Tsai2013,Soodchomshom2014,CHANG2021412552,PhysRevB.92.195423,Niu_2015,LI2017284,AN2015723,PhysRevB.103.155431,PhysRevB.105.115408}, including periodic or superlattice potentials~\cite{CHANG2021412552,PhysRevB.92.195423,Niu_2015,LI2017284,AN2015723,PhysRevB.103.155431,PhysRevB.105.115408}. 
Of particular interest are magnetic silicene superlattices (MSSLs) due to its possibilities as versatile structures with spin-valley polarization and tunneling magnetoresistance capabilities \cite{AN2015723,PhysRevB.103.155431,PhysRevB.105.115408}.  In fact, zigzag silicene nanoribbons modulated by magnetic superlattices result in valley-resolved minigaps and minibands which give rise to periodic spin-valley polarization and TMR \cite{AN2015723}.  MSSLs with structural asymmetry induced in the widths of the barriers-wells resulted in differentiated oscillating transport properties for the spin-valley components as well as the magnetization configurations~\cite{PhysRevB.103.155431}.  In specific, by adjusting the silicene's local bandgap a conductance gap in the AM configuration was found,  resulting in an enchantment of the TMR with respect to single magnetic junctions. In addition, effective positive and negative spin-valley polarization states accessible by reversing the magnetization direction were found. However, the spin-valley polarization of the AM configuration presents multiple oscillations that impede stable negative polarization states.  MSSLs with structural disorder in the widths and heights of the barriers-wells improve the spin-valley polarization and the TMR by eliminating the conductance oscillations caused by the periodic magnetic modulation~\cite{PhysRevB.105.115408}. Unfortunately, the control of the structural disorder could be tricky, and other possibilities to eliminate the conductance oscillations in MSSLs are welcomed. 

In this context, aperiodicity represents an alternative to improve the spin-valley polarization and TMR of MSSLs. In fact, aperiodicity has been instrumental to modulate/improve the spin-valley polarization and TMR of superlattices based on 2D materials~\cite{doi:10.1063/1.4827380,DeLiu_2014,LI201718}.  For instance,  in Thue-Morse graphene superlattices, the spin transport and TMR are better than in periodic graphene superlattices~\cite{doi:10.1063/1.4827380}. In fractal or Cantor graphene superlattices, the spin conductance and TMR exhibit stronger and more irregular oscillations with the Fermi energy~\cite{DeLiu_2014}.  In Thue-Morse bilayer graphene superlattices was found that the spin polarization shows drastic oscillations with the Fermi energy and the TMR presents large values that can be increased with the bias voltage~\cite{LI201718}.  As we have documented there is some progress in understanding the effects of aperiodicity on the spin-valley polarization and magnetoresistive properties of 2D material superlattices.  However, to the best of our knowledge, there are no reports studying the spin-valley polarization and TMR in aperiodic MSSLs. Taking into account the relevance of MSSLs as versatile structures we consider that a thorough assessment of the impact of aperiodicity on the spin-valley polarization and TMR is necessary. 

In this work, we study the spin-valley polarization and TMR in aperiodic MSSLs. These structures are generated by placing ferromagnetic electrodes on a silicene layer following the Fibonacci and Thue-Morse sequences. Using the Landauer-Büttiker formalism and the transfer matrix method, we investigate the transmission, electronic transport, spin-valley polarization and TMR.  Fibonacci (F-) and Thue-Morse (TM-) MSSLs are compared with its periodic counterparts. 
 
The paper is organized as follows: In Section II, we present the theoretical formalism and the expressions for the spin-valley polarization and TMR. The numerical results and discussion of the Fibonacci and Thue-Morse superlattices are given in Section III. Finally, the concluding remarks are presented in Section IV.  
 
\section{Theoretical formalism}

\begin{figure}
\includegraphics[width=0.48\textwidth]{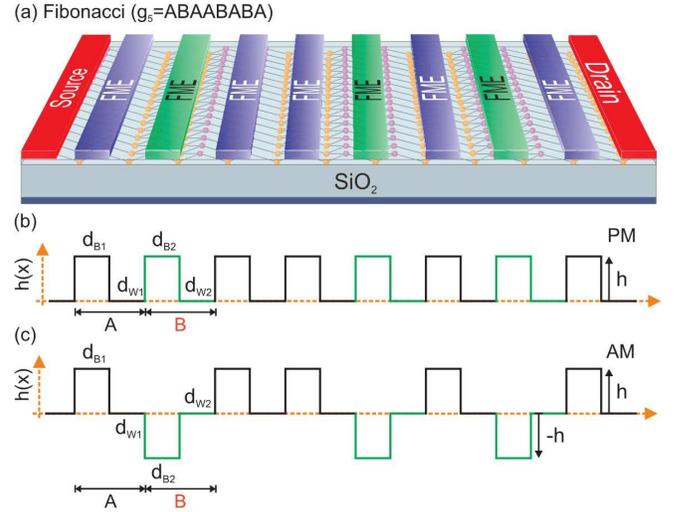}
\caption{\label{superredfibonacci} (a) Schematic representation for the fifth generation (${g}_{5}$) of Fibonacci  magnetic silicene superlattices
(F-MSSLs). The silicene layer is deposited on a SiO$_{2}$ substrate and an arrangement of ferromagnetic electrodes is placed over the silicene layer following the Fibonacci sequence. Profile of the exchange field along the superlattice axis for the (b) PM and (c) AM. There are two seeds $A$ and $B$ for the generation of aperiodic structures, each one with a FME (barrier) and a free region (well). The $A$ seed has a fixed magnetization orientation, while the $B$ seed can switch magnetization from PM  to AM.}
\end{figure}

\begin{figure}
\includegraphics[width=0.48\textwidth]{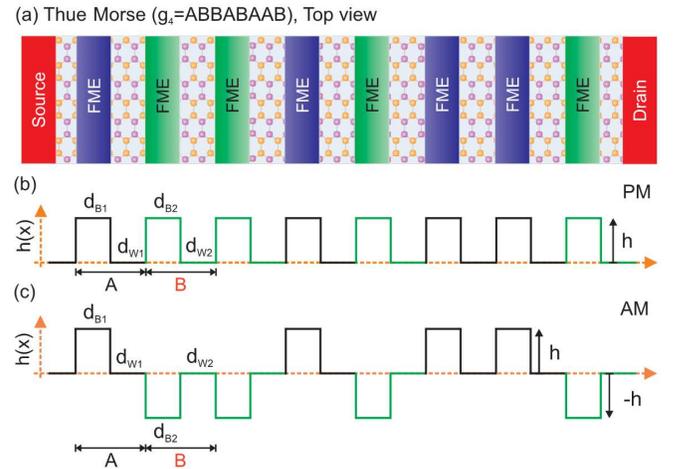}
\caption{\label{superredthuemorse} The same as Fig. \ref{superredfibonacci}, however here the aperiodic superlattice structure is obtained following the Thue-Morse sequence construction rules. The fourth generation of TM-MSSLs has been considered for the (a) schematic of the possible aperiodic superlattice structure as well as for the exchange field profile of (b) PM and (c) AM. }
\end{figure}

\begin{figure}
\begin{center}
\includegraphics[width=0.35\textwidth]{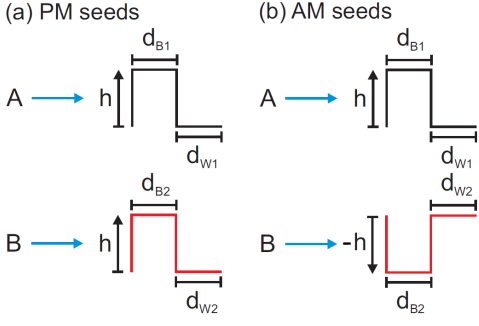}
\caption{\label{seeds}
Magnetic profiles of the $A$ and $B$ seeds for (a) PM and (b) AM. The magnetization in the $B$ seed can switch from PM to AM by reversing the magnetization direction.}
\end{center}
\end{figure}

A possible experimental configuration of the system under study is shown in Figs.~\ref{superredfibonacci}(a) and \ref{superredthuemorse}(a). It consists of a silicene layer deposited on an insulating substrate (SiO$_{2}$). An arrangement of aperiodic ferromagnetic electrodes (FMEs) are placed on the silicene layer, giving rise to the so-called aperiodic MSSLs. The FMEs with fixed and variable magnetization direction are represented with blue and green colors, respectively. Source and drain contacts are placed at the left and right side of the superlattice structure. It is important to mention that in this kind of system, the silicene layer is covered with a dielectric slab due to its high degradation in contact with the environment~\cite{Himani2021}. The aperiodic arrangement of FMEs can be generated following the Fibonacci or Thue-Morse sequence, resulting in F- or TM-MSSLs, respectively. These aperiodic superlattices can be obtained with the help of two different seeds $A$ and $B$. Each one is constituted by a FME (barrier) and a free region (well), see Fig.~\ref{seeds}.  In the case of the $A$ seed, the width of the barrier and well is denoted by $d_{B_{1}}$ and $d_{W_{1}}$, respectively. In similar fashion for the $B$ seed, the width of the barrier and well is represented by $d_{B_{2}}$ and $d_{W_{2}}$, respectively. Additionally, the $A$ seed has a fixed magnetization, while the $B$ seed can switch the magnetization orientation from PM to AM by reversing the magnetization direction, see Fig.~\ref{seeds}(a)-(b). Due to the magnetic proximity effect between the silicene layer and the FMEs that form the F- and TM-MSSLs, an exchange field of strength $h$ is generated. This field is always positive in the $A$ seed (${\theta}_{A}=1$), while in the $B$ seed, it varies depending on the magnetization configuration PM or AM (${\theta}_{B}=1$ or ${\theta}_{B}=-1$).  

F-MSSLs are generated following the Fibonacci sequence substitution rules $A\to AB$ and $B \to A$. The first and second Fibonacci sequences or generations are ${g}_{1}=A$ and ${g}_{2}=AB$, respectively. Generalizing, we have the recurrence relation ${g}_{n}={g}_{n-1}{g}_{n-2}$  for $n > 2 $,  where $n$ is the generation number, and ${g}_{n}$ denotes the $n$-th generation~\cite{PhysRevB.41.5578}. For example, the fifth generation is ${g}_{5}={g}_{4}{g}_{3 }=\left \{ ABAAB \right \}\left \{ ABA \right \}=ABAABABA$. The corresponding exchange field profile for PM and AM is shown in Fig.~\ref{superredfibonacci}(b) and \ref{superredfibonacci}(c), respectively.  On the other hand, TM-MSSLs are generated following the Thue-Morse sequence substitution rules $A\rightarrow AB$ and $B\rightarrow BA$. The first and second Thue-Morse generations are defined as ${g}_{1}=A$ and ${g}_{2}=AB$, respectively. Generalizing, the recurrence relation takes the form ${g}_{n}={g}_{n-1}\overline{g}_{n-1}$ for $n \geq 3$, where $n$ is the generation number, ${g}_{n}$ denotes the $n$-th sequence and $\overline{g}_{n}$ is the complement of ${g}_{n}$ obtained by interchanging $A$ and $B$ in ${g}_{n}$~\cite{Liu_1998}. Then, the fourth generation is ${g}_{4}={g}_{3}\overline{g}_{3}=\left \{ ABBA \right \}\left \{ BAAB\right \}=ABBABAAB$. The corresponding exchange field profile for PM and AM is shown in Fig.~\ref{superredthuemorse}(b) and \ref{superredthuemorse}(c), respectively. 

To describe the charge carriers through F- and TM-MSSLs, we employ a low-energy effective Hamiltonian~\cite{Wang_2014,wangh2019,PhysRevLett.107.076802,PhysRevB.84.195430,PhysRevB.85.075423,PhysRevB.103.155431},
\begin{equation}\label{hamiltonianosiliceno}
\hat{H}={v}_{F}({p}_{x}{\sigma}_{x}-{\tau}_{z}{p}_{y}{ \sigma}_{y})-({s}_{z}{\tau}_{z}{\Delta}_{SO}-{\Delta}_{z}(x)) {\sigma}_{z}-{s}_{z}h(x),
\end{equation}

\noindent{where}
\begin{equation}
{\Delta}_{z}(x) =\left\{\begin{matrix}
{\Delta}_{z} & \text{for barriers and wells}\\ 
 0 & \text{otherwise}, 
\end{matrix}\right.
\end{equation}

\noindent{and} 
\begin{equation}\label{h}
h(x) =\left\{\begin{matrix}
{\theta}_{A,B} h & \text{for barriers}\\ 
0 &  \text{for  wells}.
\end{matrix}\right.
\end{equation}

\noindent Here, $\vec{p}=\left ( {p}_{x},{p}_{y} \right )$ and $\vec{\sigma}=\left ( {\sigma}_{x},{\sigma}_{y},{\sigma}_{z} \right )$ are the two-dimensional momentum and the vector of Pauli matrices related to the sublattice pseudospin, respectively. ${\Delta}_{SO}=3.9$ meV is the spin-orbit interaction, ${v}_{F}\approx 0.5 \times {10}^{6} \ m/s$ is the Fermi velocity of the charge carriers in silicene, ${\tau}_{z}=\pm 1$ and ${s}_{z}=\pm 1$ stand for valley and spin indices.  
$\Delta_{z}$ represents the silicene on-site potential or local band gap that can be modulated through an electric field applied perpendicularly to the silicene through the FMEs.

To obtain the transmission probability of an electron through F- and TM-MSSLs, we use the transfer matrix method~\cite{Soukoulis2008}. The transfer matrix ${M}_{{g}_{n}}^{SL}$ of the $n$-th generation relates the coefficients of the wave functions in the left semi-infinite region with the corresponding ones in the right semi-infinite region, this relationship can be written as
\begin{equation}\label{matrizdetransferencia} 
\binom{A_{+}^{L}}{A_{-}^{L}}= {M}_{{g}_{n}}^{SL} \binom{A_{+}^{R}}{A_{-}^{R}}.
\end{equation}

\noindent The transfer matrix (${M}_{{g}_{n}}^{SL}$) has all the information of the wave function in each barrier and well of the superlattice structure. The wave functions in the barrier regions are given by
\begin{equation}\label{funcion de onda}
\Psi_{\pm}^{b}(x,y)=A_{\pm}^{b}
\left( \begin{matrix} 1 \\ 
v_{\pm}^{b} \end{matrix} \right) 
e^{\pm i{q}_{x}^{b}x+i{q}_{y}^{b}y},
\end{equation}

\noindent{where }
\begin{equation}\label{constante.v}
{v}_{\pm}^{b}=\frac{{v}_{F} \hbar ({\pm q}_{x}^b-i{\tau}_{z}{q}_{y}^b)}{E+{s}_{z} \theta_{A,B} h - ({s}_{z} {\tau}_{z} {\Delta}_{SO}-{\Delta}_{z})},
\end{equation}

\noindent{and }
\begin{equation}\label{constante.q}
{q}_{x}^{b}=\frac {1}{{v}_{F}\hbar} \sqrt{{(E+{s}_{z}\theta_{A,B} h)}^{2}-{({\tau}_{z}{s}_{z} {\Delta}_{SO}-{\Delta}_{z})}^{2}-{(\hbar {v}_{F} {q}_{y}^{b})}^2}.
\end{equation}

\noindent For the wells regions, the wave functions and wave vectors are obtained by setting $h=0$ and changing the superscript $b\rightarrow w$ in Eqs. (\ref{funcion de onda}), (\ref{constante.v}), and (\ref{constante.q}).

The transfer matrix ${M}_{{g}_{n}}^{F-SL}$ for the $n$-th generation of F-MSSLs can be written as the product of the transfer matrices for the $n-1$ and $n-2$ generations, which is in agreement with the recurrence relation for the Fibonacci sequence, namely:
\begin{equation}\label{matrizdetransferenciaF} 
{M}_{{g}_{n}}^{F-SL}={M}_{{g}_{n-1}}^{F-SL} {M}_{{g}_{n-2}}^{F-SL}.
\end{equation}

\noindent{The above is true for ${g}_{n} \geq {g}_{3}$, with ${M}_{{g}_{1}}^{F-SL}={M}_{A}$ and ${M}_{{g}_{2}}^{F-SL}={M}_{A}{M}_{B}$.}

Also, the transfer matrix ${M}_{{g}_{n}}^{TM-SL}$ for the $n$-th generation of TM-MSSLs can be written as the product of the transfer matrix and its complement matrix for the $n-1$ generation, which is in agreement with the recurrence relation for the Thue-Morse sequence, namely:
\begin{equation}\label{matrizdetransferenciaTM} 
{M}_{{g}_{n}}^{TM-SL}={M}_{{g}_{n-1}}^{TM-SL} \overline{M}_{{g}_{n-1}}^{TM-SL}.
\end{equation}

\noindent{This relation is valid for ${g}_{n}\geq{g}_{3}$, where ${M}_{{g}_{1}}^{TM-SL}={M}_{A}$, ${M}_{{g}_{2}}^{TM-SL}={M}_{A}{M}_{B}$ and ${M}_{{g}_{3}}^{TM-SL}={M}_{A}{M}_{B}{M}_{B}{M}_{A}$. The complement $ \overline{M}_{{g}_{n-1}}^{TM-SL}$ is obtained by interchanging ${M}_{A}$ and ${M}_{B}$ in ${M}_{{g}_{n-1}}^{TM-SL}$.}

The transfer matrices $ {M}_{{g}_{n}}^{F-SL}$ and ${M}_{{g}_{n}}^{TM-SL}$ are written in terms of the transfer matrices ${M}_{A}$ and ${M}_{B}$. The matrices ${M}_{A}$ and ${M}_{B}$ correspond to the seeds $A$ and $B$, respectively. These matrices are expressed in terms of the transfer matrices associated with the barriers and wells in each region,
\begin{eqnarray}
{M}_{A} &=& {M}_{{b}_{1}} {M}_{{w}_{1}},\\
{M}_{B} &=& {M}_{{b}_{2}} {M}_{{w}_{2}},
\end{eqnarray}

\noindent{where}
\begin{eqnarray}
{M}_{{b}_{1}} & =& {D}_{0}^{-1} ({D}_{{b}_{1}} {P}_{{b}_{1}} {D}_{{b}_{1}}^{-1}) {D}_{0},\\
{M}_{{w}_{1}} & =& {D}_{0}^{-1} ({D}_{{w}_{1}} {P}_{{w}_{1}} {D}_{{w}_{1}}^{-1}) {D}_{0},\\
{M}_{{b}_{2}} & =& {D}_{0}^{-1} ({D}_{{b}_{2}} {P}_{{b}_{2}} {D}_{{b}_{2}}^{-1}) {D}_{0},\\
{M}_{{w}_{2}} & =& {D}_{0}^{-1} ({D}_{{w}_{2}} {P}_{{w}_{2}} {D}_{{w}_{2}}^{-1}) {D}_{0},
\end{eqnarray}

\noindent{here ${D}_{{b}_{i}}$ and ${D}_{{w}_{i}}$ (${P}_{{b}_{i}}$ and ${P}_{{w}_{i}}$) are the dynamic (propagation) matrices in the barrier and well regions, respectively.  ${D}_{0}$ corresponds to the dynamic matrix of the semi-infinite 
regions. These matrices are given as}
\begin{equation}\label{matrizdetransferenciab1} 
{ D }_{ i }=\begin{pmatrix} 1 & 1 \\ { v }_{ + }^{ i } & { v }_{ - }^{ i } \end{pmatrix},\quad
\end{equation}

\noindent{and}
\begin{equation}\label{matrizdetransferenciab1} 
{ P }_{ i }=\begin{pmatrix} { e }^{ -i{ q }_{ x }^{ i } {d}_{i} } & 0 \\ 0 & { e }^{ i{ q }_{ x }^{ i } {d}_{i} }, \end{pmatrix}
\end{equation}

\noindent{where $i=0,{b}_{1}, {b}_{2}, {w}_{1}, {w}_{2}$ runs over the semi-infinite, barrier and well regions.}

By definition, the transmittance is given in terms of the probability density fluxes ${j}_{x}^{out}$ and ${j}_{x}^{in}$ associated with the outgoing waves in the right semi-infinite region and to the incoming waves in the left semi-infinite region, respectively. Specifically, the transmission is given by
\begin{equation}\label{densityfluxes}
{T}_{{\tau}_{z} {s}_{z}}^{{\theta}_{A} {\theta}_{B}}={\left | \frac{{j}_{x}^{out}}{{j}_{x}^{in}} \right |},
\end{equation}
where ${j}_{x}={v}_ {F} {\psi}^{\dagger} {\sigma}_{x} \psi$ is the $x$-component of the probability density flux of electrons. Then, we can rewrite the transmission in terms of the amplitude of the outgoing and incoming wave functions, 
\begin{equation}\label{transmision3}
{T}_{{\tau}_{z} {s}_{z}}^{{\theta}_{A} {\theta}_{B}}= \frac { { q }_{x}^{R} }{ { q }_{ x }^{L} }  \frac {[ (E+{ s }_{ z }{ h }_{ L }{ \phi  }_{ L })-( {\tau}_{z} {s}_{z} {\Delta}_{SO} -  { \Delta  }_{ z }^{L})] }{[ (E+{ s }_{ z }{ h }_{ R }{ \phi  }_{ R })-( {\tau}_{z} {s}_{z} {\Delta}_{SO} -  { \Delta  }_{ z }^{R})] } {\left | \frac{{A}_{+}^{R}}{{A}_{+}^{L}} \right |}^{2},
\end{equation} 

\noindent where $q^L_x$, $h_L$, $\Delta^L_z$ and $\phi_L$ ($q^R_x$, $h_R$, $\Delta^R_z$ and $\phi_LR$) are the wave vector, exchange field strength, on-site potential and angle of incidence of the incoming (outgoing) wave in the left (right) semi-infinite region, respectively. 

\noindent In our case, the semi-infinite regions are the same, as a consequence its underlying physical properties, then the transmission adopts the following form:
\begin{equation}\label{transmision3}
{T}_{{\tau}_{z} {s}_{z}}^{{\theta}_{B} {\theta}_{A}}= {\left | \frac{{A}_{+}^{R}}{{A}_{+}^{L}} \right |}^{2}={\left | \frac{1}{M_{11}^{SL}} \right |}^{2},
\end{equation}

\noindent{where $M_{11}^{SL}$ is the (1,1) element of the transfer matrix ${M}_{{g}_{n}}^{SL}$.}

According to the Landauer-Büttiker formalism~\cite{datta1995}, the spin-valley conductance for a particular magnetization configurations at zero temperature is given by
\begin{equation}\label{conductancia}
{ G }_{{\tau}_{z} {s}_{z}}^{ {\theta}_{A} {\theta}_{B} }={ G }_{ 0 }\int _{ -\frac { \pi  }{ 2 }  }^{ \frac { \pi  }{ 2 }  }{ { T }_{{\tau}_{z} {s}_{z}}^{ {\theta}_{A} {\theta}_{B} }}  \cos \phi d\phi,
\end{equation}

\noindent{where $\phi$ is the angle of the impinging charge carries, ${G}_{0}={e}^{2} {L}_{y} {k}_{F}/(2 \pi h )$ is the fundamental conductance factor, ${L}_{y}$ represents the width of the silicene sheet, and ${k}_{F}=\sqrt{{E}^{2}-{\Delta}_{SO}^{2}}/ \hbar {v}_{F}$ the Fermi wave vector in the semi-infinite free regions.}

The spin-valley polarizations for the magnetization configurations are given as~\cite{doi:10.1063/1.3473725,doi:10.1063/1.3569621},
\begin{equation}\label{polarizacionsz}
\eta_{s}^{ {\theta}_{A} {\theta}_{B}}=\frac { \sum _{ {\tau}_{z}  }^{  }{ { G }_{ \tau_z ,\uparrow  }^{ {\theta}_{A} {\theta}_{B}  }-{ G }_{ \tau_z ,\downarrow  }^{ {\theta}_{A} {\theta}_{B}   } }  }{ \sum _{ {\tau}_{z} {s}_{z}  }^{  }{ { G }_{ {\tau}_{z} {s}_{z}   }^{ {\theta}_{A} {\theta}_{B}  } }  },
\end{equation}

\noindent{and}
\begin{equation}\label{polarizaciontz}
\eta_{ v }^{ {\theta}_{A} {\theta}_{B}  }=\frac { \sum _{ {s}_{z}  }^{  }{ { G }_{ K {s}_{z}  }^{ {\theta}_{A} {\theta}_{B}  }-{ G }_{ {K}^{'} {s}_{z}  }^{ {\theta}_{A} {\theta}_{B}   } }  }{ \sum _{ {\tau}_{z} {s}_{z}  }^{  }{ { G }_{ {\tau}_{z} {s}_{z}   }^{ {\theta}_{A} {\theta}_{B}  } }  }.
\end{equation}

Finally, the TMR is defined as~\cite{PhysRevB.92.245412}, 
\begin{equation}\label{saxenaTMR}
{\rm TMR}=\frac{G_{c}^{\Uparrow \Uparrow} - G_{c}^{\Uparrow \Downarrow}}{G_{c}^{\Uparrow \Uparrow}},
\end{equation}

\noindent{here $G_{c}^{\Uparrow \Uparrow}=\sum_{{\tau}_{z} {s}_{z}}^{  }{{G}_{{\tau}_{z}{s}_{z}}^{\Uparrow \Uparrow}}$ $(G_{c}^{\Uparrow \Downarrow}=\sum_{{\tau}_{z} {s}_{z}}^{  }{{G}_{{\tau}_{z}{s}_{z}}^{\Uparrow \Downarrow}})$ is the total charge conductance for the PM (AM) configuration.

\section{Results and discussion}

\begin{figure*}[htb!]
\begin{center}
\includegraphics[width=0.9\textwidth]{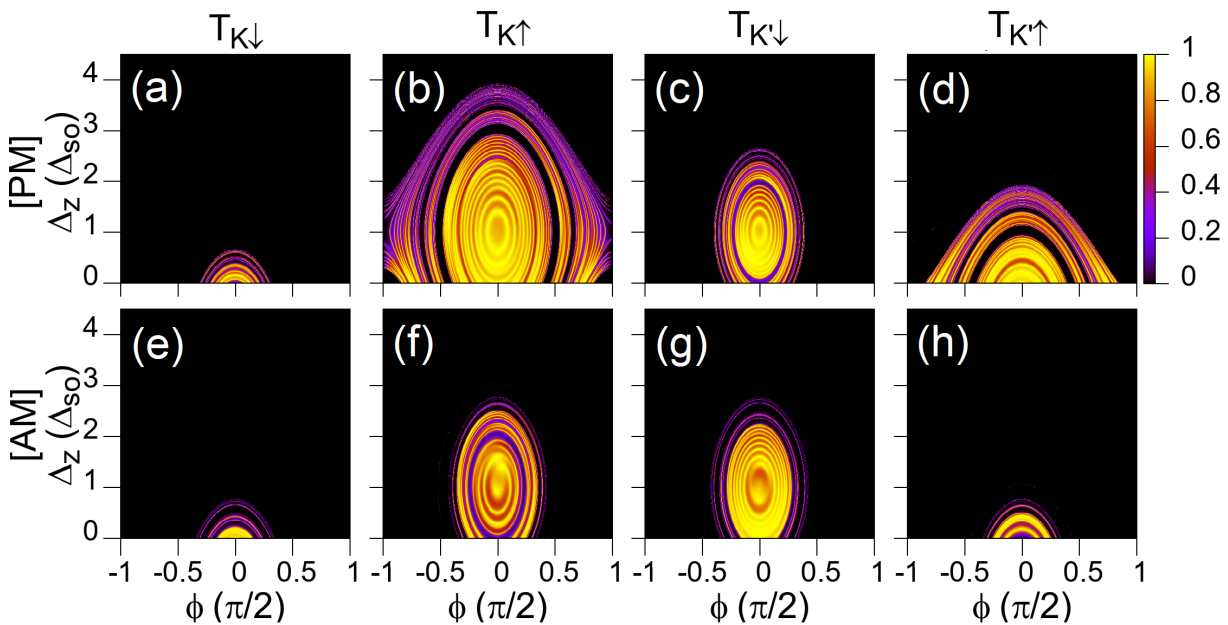}
\caption{\label{7fac} Transmission maps as a function of $\Delta_{z}$ and $\phi$ for the seventh generation (${g}_{7}$) of F-MSSLs. (a)-(d) PM and (e)-(h) AM configuration for the different spin-valley components.}
\end{center}
\end{figure*}

In the present section, we analyze the numerical results of the transport properties of F- and TM-MSSLs. For all results shown below, the widths considered for the barriers and wells that constitute the seeds $A$ and $B$ are ${d}_{{B}_{1}}= 2 {l}_{SO}$, ${d}_{{B}_{2}}=3 {l}_{SO}$, ${d}_{{W}_{1}}= 1 {l}_{SO}$, and ${d}_{{W}_{2}}= 2 {l}_{SO}$. While, the exchange field in the barriers and the energy of the charge carriers are $h = 1.2 {\Delta}_{SO}$ and $E=3.0{\Delta}_{SO}$, respectively. The widths and energies are given in units of ${l}_{SO}=\hbar {v}_{F}/{\Delta}_{SO}=89.6$ nm, and ${\Delta}_{SO}$, respectively. We focus on the variation of ${\Delta}_{z}$ as a practical modulating parameter. All cases for F- and TM-MSSLs are compared with  periodic MSSLs (P-MSSLs), and all superlattices have structural asymmetry, i.e., superlattices in which the structural parameters are different from each other. The number of superlattice periods is denoted by $NP_{n}$, where $n$ is the number of unit-cell repetitions. The periodic unit-cell is composed by two barriers ($B$) and two wells ($W$). The first superlattice periods are $NP_{1}=\{BWBW\}$, $NP_{2}=\{BWBW\}\{BWBW\}$, $NP_{3}=\{BWBW\}\{BWBW\}\{BWBW\}$, and so on. 

\subsection{Fibonacci MSSLs}
The transport properties of charge carriers through F-MSSLs are analyzed in this section. In Fig.~\ref{7fac}, we show the transmission maps as a function of ${\Delta}_{z}$ and $\phi$ for the seventh Fibonacci generation (${g}_{7}$). The first row of panels corresponds to the PM configuration, while the second row corresponds to the AM configuration. We identify PM when $\theta_{A}\theta_{B}\to \Uparrow \Uparrow$, and AM when $\theta_{A}\theta_{B}\to \Uparrow \Downarrow$. The same for the valleys when $\tau_{z}=+1 (-1) \to K (K^{'})$. As we can see, all transmission maps for PM and AM are different for all spin-valley components. We can also see semicircular regions with high and low transmission probabilities as a consequence of the aperiodic magnetic modulation. On the other hand, for the AM case, the highest transport probability is concentrated between $-0.5\;(\pi/2)< \phi < +0.5\;(\pi/ 2)$ in all the transmission maps. Furthermore, in the PM case, the components $T_{K_{\downarrow}}$ and $T_{K'_{\downarrow}}$ have practically zero transmission probability for angles greater than $\pm \;0.5\;(\pi/2)$, while for the $T_{K_{\uparrow}}$ and $T_{K'_{\uparrow}}$ components it is possible to have higher transmission probabilities for angles $|\phi|\leq \pi/2$. All these characteristics of the transmission maps will be reflected in the spin-valley polarization and magnetoresistance.

\begin{figure}[htb!]
\begin{center}
\includegraphics[width=0.48\textwidth]{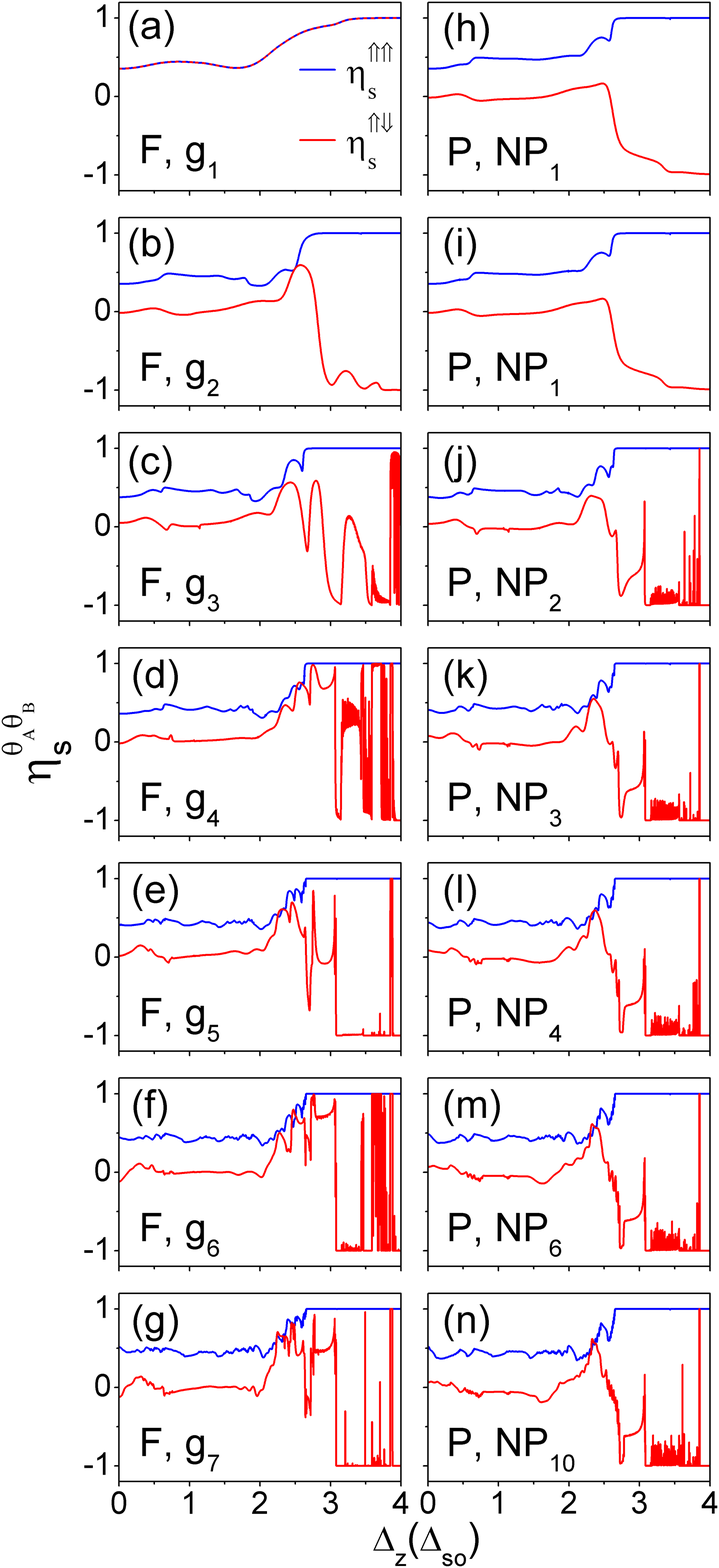}
\caption{\label{fapsz} Spin polarization as a function of $\Delta_{z}$ for (a)-(g) F-
and (h)-(n) P-MSSLs. The blue and red lines correspond to the PM and AM configurations, respectively.}
\end{center}
\end{figure}

\begin{figure}[htb!]
\begin{center}
\includegraphics[width=0.48\textwidth]{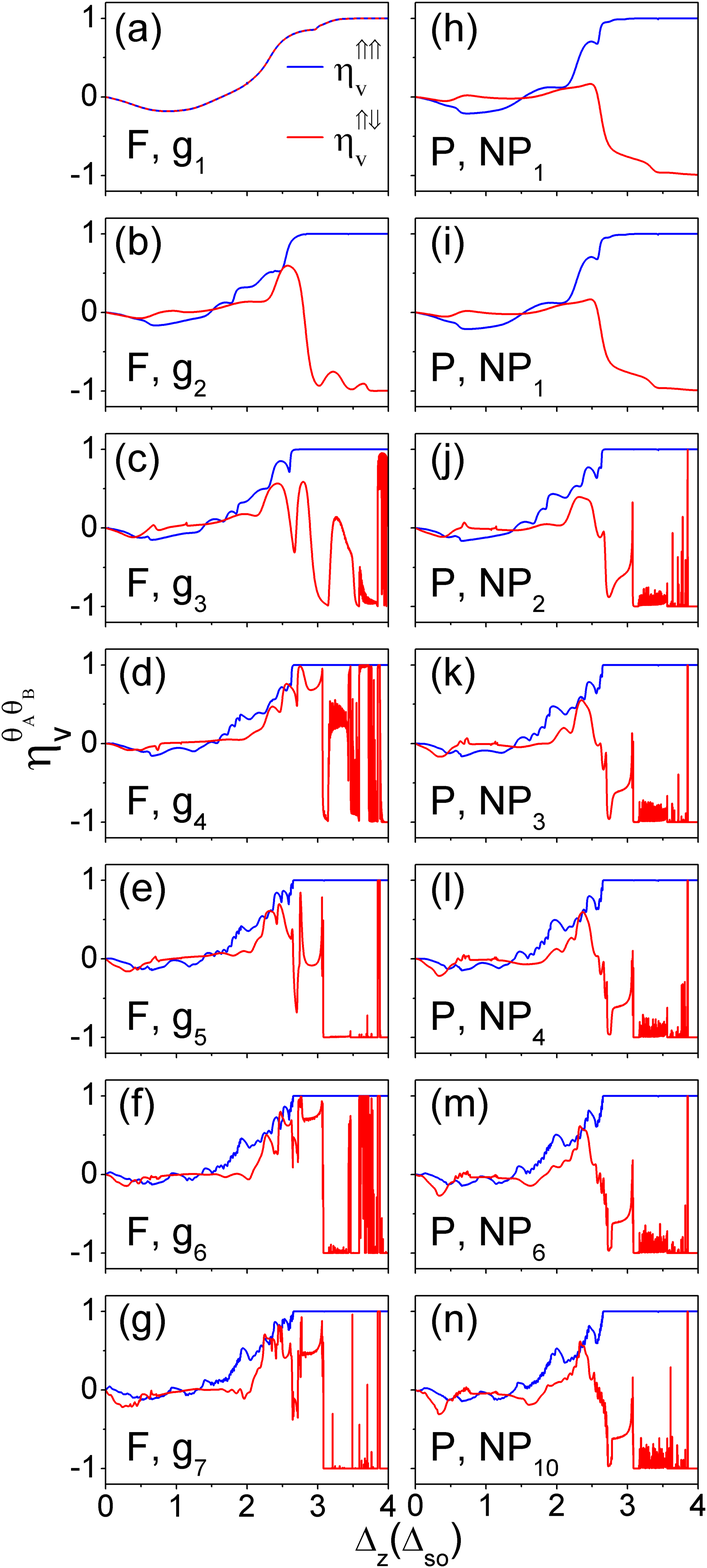}
\caption{\label{fapst} Valley polarization as a function of $\Delta_{z}$ for (a)-(g)  F- and 
(h)-(n) P-MSSLs. The blue lines and red lines correspond to the PM and AM configurations, respectively.}
\end{center}
\end{figure}

The spin polarization as a function of ${\Delta}_{z}$ is shown in Fig.~\ref{fapsz}. We analyze the impact of F-MSSLs on the spin polarization and compare it with P-MSSLs as reference. For this comparison, we have considered that the difference in the number of barriers between F- and P-MSSLs does not surpass one barrier, so they have similar sizes. In F-MSSLs, we can see that for the first generation, the polarization for the PM (blue line) and AM (red line) configurations are overlapping because there is only one barrier in the system. From the second generation, the PM configuration reaches almost complete polarization after $\Delta_{z}=2.7 \Delta_{SO}$, being better defined as the generation increases, see Fig.~\ref{fapsz}(a)-(g). On the contrary, for the AM configuration, the spin polarization shows oscillations that increase as the generation increases. Notice that in the even generations the spin polarization has more oscillations, while the odd generations have fewer oscillations, producing windows with 100\% spin polarization. This difference is because in the even generations the last barrier has negative magnetization, while in the odd ones the last barrier has positive magnetization. This, in principle, subtle difference brings considerable changes in the constructive-destructive interference phenomena of the AM configuration, reducing the oscillations of the spin polarization in the odd generations.
Regarding P-MSSLs, the PM configuration reaches complete polarization after $\Delta_{z}=2.7 \Delta_{SO}$ in practically all superlattice periods. In addition, the region of perfect polarization is better defined as the number of periods increases.  In the AM configuration, many oscillations are observed in all generations.  In comparison with the AM configuration of F-MSSLs, here the spin polarization shows plenty of oscillations regardless the parity of the superlattice structure, see Fig.~\ref{fapsz}(h)-(n).

Figure~\ref{fapst} shows the valley polarization as a function of ${\Delta}_{z}$ for F- and P-MSSLs. In the first generation of F-MSSLs, the valley polarization of PM (blue line) and AM (red line) configurations are overlapping. Even more, in the range $0<{\Delta}_{z}<2{\Delta}_{SO}$ the valley polarization is negative, after that the polarization increases, reaching 100\% polarization. From the second to the seventh generation the PM configuration reaches almost complete positive polarization after $\Delta_{z}\approx 2.6\;\Delta_{SO}$, being better defined as the generation increases, see Fig.~\ref{fapst}(a)-(g). While, for the AM configuration, the valley polarization is close to zero up to $\Delta_{z}\approx 2.0 \Delta_{SO}$, after that it has oscillations that make it going from positive to negative polarization values. Here, it is also important to mention that as in the case of the spin polarization, the valley polarization is highly dependent on the parity of generations, see Fig. \ref{fapst}(a)-(g). 
In P-MSSLs, the valley polarization gets 100\% positive values after $\Delta_{z}\approx 2.7 \Delta_{SO}$ for the PM configuration.  For the AM configuration, the valley polarization tends to be negative, with a greater number of oscillations as the generations increases, see Fig.~\ref{fapst}(h)-(n). In similar fashion as for the spin polarization, the valley polarization of P-MSSLs has plenty of oscillations in the AM configuration, see Fig. \ref{fapst}(h)-(n). 

Finally, the TMR as a function of ${\Delta}_{z}$ for F- (green line) and P-MSSLs (black line) is shown in Fig.~\ref{fatmr}. In both systems, there are oscillations whose amplitude grows as the generation increases. There are also TMR values of 100\% for ${\Delta}_{z}> 2.8 \Delta_{SO}$ in both systems. This high polarization windows are presented in practically all generations (periods), and as the generation (period) increases they are better defined. If we focus on the range $2.5{\Delta}_{SO}<{\Delta}_{z}<2.8{\Delta}_{SO}$, the peaks corresponding to P-MSSLs are greater than those of the Fibonacci case. However, from the fourth generation onwards the TMR values are better for F-MSSLs. In particular, for the sixth and seventh generations, the TMR shows an enhancement of about 35\% for F-MSSLs with respect to P-MSSLs.
\begin{figure}
\begin{center}
\includegraphics[width=0.49\textwidth]{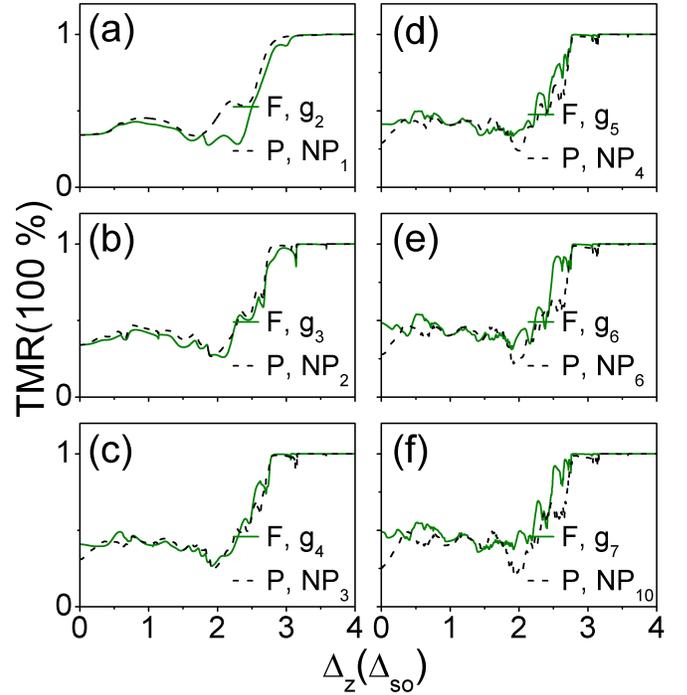}
\caption{\label{fatmr} TMR as a function of $\Delta_z$ for (a)-(c) F- and (d)-(f) P-MSSLs, green and black curves, respectively.}
\end{center}
\end{figure}

\subsection{Thue-Morse MSSLs}

Now, we analyze the transport properties of the charge carriers through TM-MSSLs. Fig.~\ref{5tac} shows the transmission maps as a function of ${\Delta}_{z}$ and $\phi$ for the fifth TM-generation (${g}_{5}$). The first row of panels corresponds to the PM configuration, while the second row corresponds to the AM configuration. We can see that the transmission maps are different from each other for all spin-valley components. There is a more noticeable difference between the transmission maps for TM-MSSLs with respect to F-MSSLs, especially, because semicircular regions with high and low transmission arise within the bands as well as in the bandgap regions. In addition, for the AM case, we can see similarities between the transmission maps of the spin-up (spin-down) electrons in the $K$ valley and spin-down (spin-up) electrons in the ${K}^{'}$ valley. 
\begin{figure*}[htb!]
\begin{center}
\includegraphics[width=0.9\textwidth]{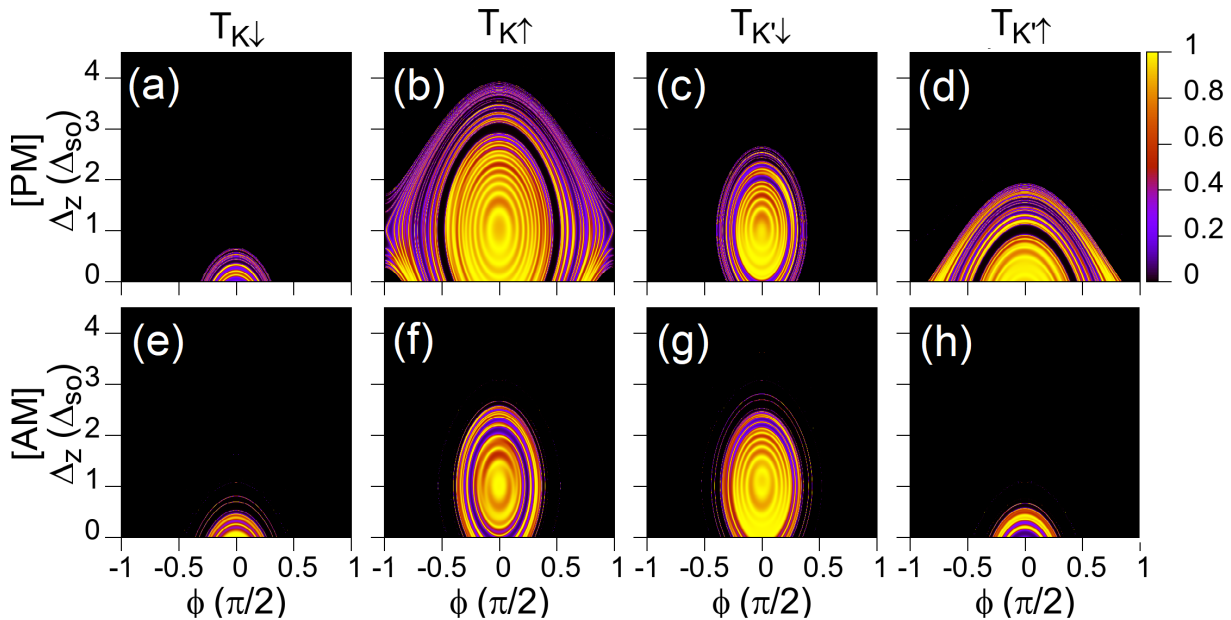}
\caption{\label{5tac}Transmission maps as a function of $\Delta_{z}$ and $\phi$ for the fifth generation 
(${g}_{5}$) of TM-MSSLs. (a)-(d) PM and (e)-(h) AM configurations for the different spin-valley components.}
\end{center}
\end{figure*}

Now, it is time to analyze the impact of Thue-Morse magnetic modulation on the spin polarization. Fig.~\ref{tapsz}(a)-(d) shows the spin polarization as a function of ${\Delta}_{z}$ for TM-MSSLs. We can observe that the PM (blue lines) configuration reaches almost complete positive polarization after $\Delta_{z}=2.7 \Delta_{SO}$ for all generations. Furthermore, for the AM configuration (red lines), there are fewer oscillations, and after the third generation wide 100\% polarization windows take place. In particular, in the fourth and fifth generation, the spin polarization is practically flat after $\Delta_{z}=3.2 \Delta_{SO}$. In the case of P-MSSLs, the spin polarization of the PM configuration shows similar behavior as for TM-MSSLs, reaching almost complete polarization after $\Delta_{z}=2.7 \Delta_{SO}$, see Fig.~\ref{tapsz}(e)-(h). In the AM configuration,  the spin polarization presents oscillations that increase with the superlattice period $NP_n$.  These oscillations impede windows of perfect spin polarization, being this the main problem with P-MSSLs as versatile structures. In short, we can see that after the third generation TM-MSSLs open the door to two well-defined spin polarization states reachable by reversing the magnetization direction. 

\begin{figure}[htb!]
\begin{center}
\includegraphics[width=0.48\textwidth]{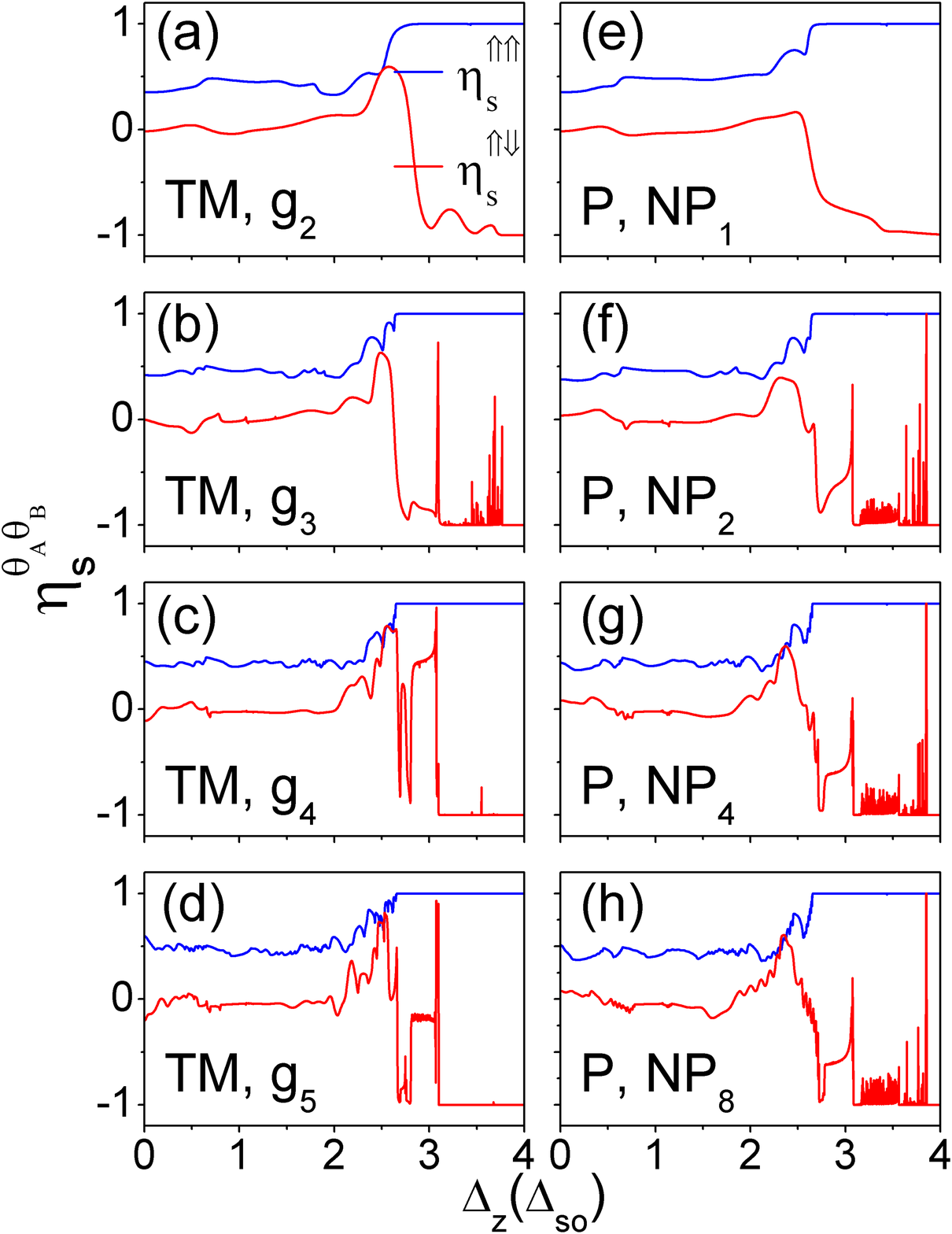}
\caption{\label{tapsz} Spin polarization as a function of $\Delta_{z}$ for (a)-(d) TM- and 
(e)-(h) P-MSSLs. The blue and red lines correspond to the PM and AM configuration, respectively.}
\end{center}
\end{figure}

\begin{figure}[htb!]
\begin{center}
\includegraphics[width=0.48\textwidth]{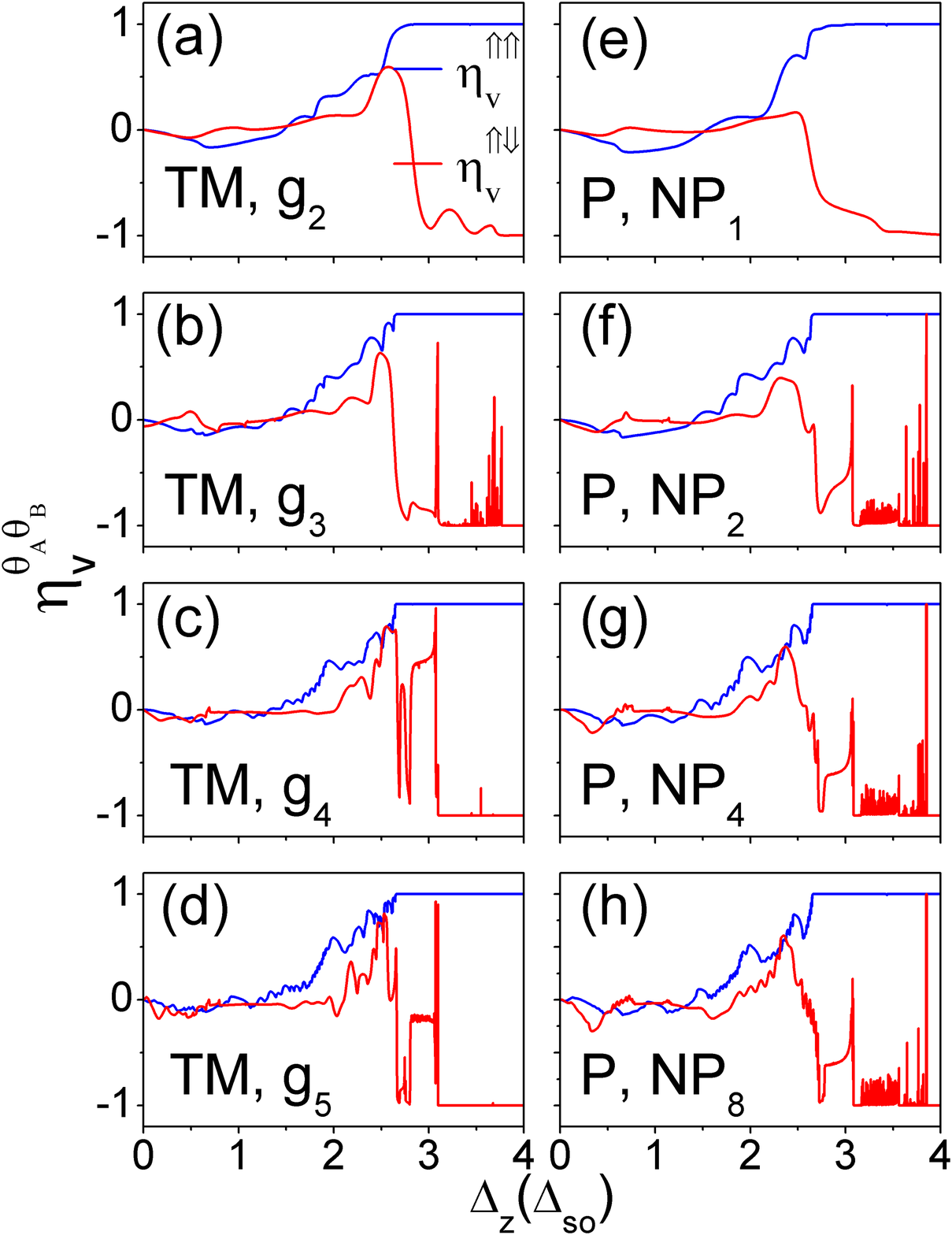}
\caption{\label{tapst}Valley polarization as a function of $\Delta_{z}$ for (a)-(d) TM- and (e)-(h) P-MSSLs. The blue and red lines correspond to the PM and AM configuration, respectively.}
\end{center}
\end{figure}

The valley polarization as a function of ${\Delta}_{z}$ for TM- and P-MSSLs is shown in Fig.~\ref{tapst}. From the second to fifth generation of TM-MSSLs, the PM configuration reaches almost complete positive polarization after $\Delta_{z}\approx 2.7 \Delta_{SO}$, being better defined as the generation increase. For the AM configuration, the valley polarization presents some oscillations in the third generation $g_{3}$. But in the $g_{4}$ and $g_{5}$ generations, in the range $3.1{\Delta}_{SO}<{\Delta}_{z}<4{\Delta}_{SO}$, the oscillations disappear and wide 100\% polarization windows arise, see Fig.~\ref{tapst}(a)-(d). For P-MSSLs, from the second to fifth period, the PM configuration reaches complete positive polarization, while in the AM configuration, the valley polarization tends to be negative with oscillations that increase as the period increases, see Fig.~\ref{tapst}(e)-(h). It is notable the difference with respect to TM-MSSLs, where the oscillations disappear. Here, it is important to remark that, as in the case of the spin polarization, TM-MSSLs also open the door to two well-defined valley polarization states reachable by reversing the magnetization direction for $g_n > g_3$. 

\begin{figure}[htb!]
\begin{center}
\includegraphics[width=0.48\textwidth]{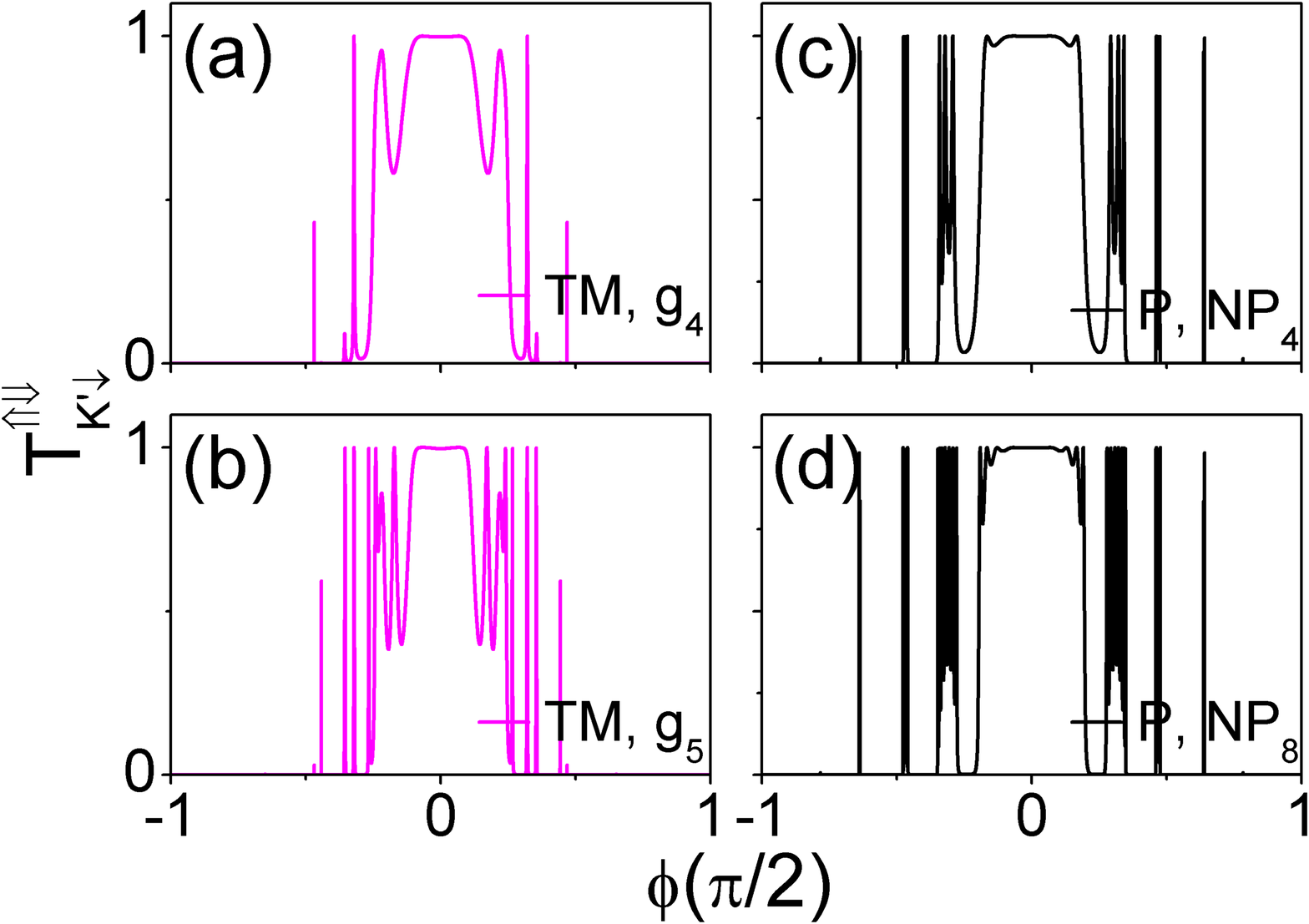}
\caption{\label{Impact-TM-Aperiodicity} Impact of TM aperiodicity on the transmission properties of MSSLs for (a) $g_4$ and (b) $g_5$.  The K' spin-down component of the transmittance for the AM configuration ($T^{\Uparrow \Downarrow}_{K'\downarrow}$) as a function of the angle of incidence ($\phi$) has been considered. The periodic counterparts (c) $NP_4$ and (d) $NP_8$ are also shown to contrast TM aperiodic effects.  Here, the on-site potential is set to $\Delta_z=0$. }
\end{center}
\end{figure}

\begin{figure}
\begin{center}
\includegraphics[width=0.48\textwidth]{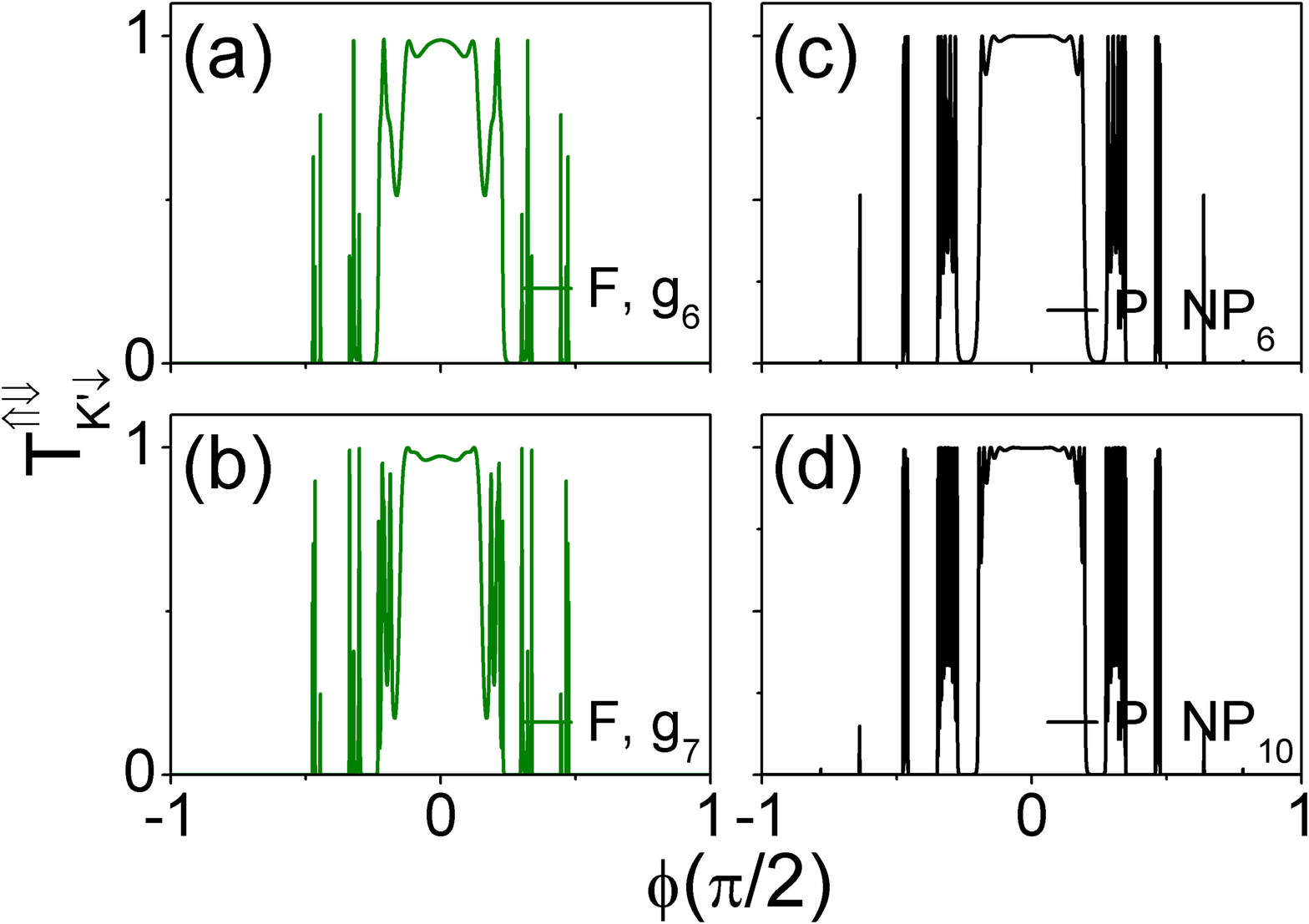}
\caption{\label{Impact-F-Aperiodicity} The same as in Fig. \ref{Impact-TM-Aperiodicity}, but for Fibonacci aperiodicity.  The superlattice generations (periods) considered are (a) $g_6$ and (b) $g_7$ ((c) $NP_6$ and (d) $NP_{10}$).}
\end{center}
\end{figure}

In order to have a better understanding of these significant differences of the spin-valley polarization of TM-MSSLs with respect to F- and P-MSSLs we will analyze in more detail the impact of TM aperiodicity on the transmission properties of MSSLs. In Fig. \ref{Impact-TM-Aperiodicity} we show the transmittance as a function of the angle of incidence for (a) $g_4$ and (b) $g_5$ of TM-MSSLs. The K$^{'}$ spin-down component of the transmission for the AM configuration ($T^{\Uparrow \Downarrow}_{K'\downarrow}$) has been considered. The on-site potential has been set to $\Delta_z=0$. The periodic counterparts (c) $NP_4$ and (d) $NP_8$ have been considered to contrast the TM aperiodicity effects.  As we can notice in the case of P-MSSLs there are well-defined minibands and minigaps. The number of resonances in the minibands increases with the number of superlattice periods.  In addition, the angular width of the minibands gets smaller as they deviate from normal incidence.  In the case of TM-MSSLs the minibands get fragmented and some resonances lie in the minigaps of P-MSSLs.  The fragmentation is more prominent as the TM generation increases.  Another characteristic, that is quite relevant for the transport properties, is the transmission gap that the TM aperiodicity creates, see the angular interval define by $\phi > 0.4 (\pi/2)$ in Fig.  \ref{Impact-TM-Aperiodicity}(a) and  \ref{Impact-TM-Aperiodicity}(b).  All these characteristics are also presented in general in the other spin-valley component of the transmittance. What is most important is that they give rise to a conductance gap that results in a window of perfect spin-valley polarization as we have documented, see Figs.  \ref{tapsz}(d) and \ref{tapst}(d).  Here, it is important to mention that in general disorder or disordered structures give rise to localized states reducing considerably the transport properties \citep{anderson1958,chomette1986}.  In this sense and according to the transmission and transport characteristics of TM-MSSLs is that they are considered as structures with higher disorder.  This contrasts with the transmission characteristics of F-MSSLs as shown in Fig.  \ref{Impact-F-Aperiodicity}(a) and \ref{Impact-F-Aperiodicity}(b). In this case,  the main effect of the aperiodicity is the fragmentation of the minibands. In fact, there are no resonances in the minigaps of the P-MSSLs and a transmission gap does not arise as in the case of TM-MSSLs, compare the transmission characteristics of F-MSSLs with the ones of P-MSSLs in Fig. \ref{Impact-F-Aperiodicity}. These transmission characteristics of F-MSSLs avoid the creation of a conductance gap, and as a consequence, perfect spin-valley polarization windows as in the case of TM-MSSLs. In this context, F-MSSLs are regarded as superlattices closer to ordered structures.  In short, aperiodic MSSLs with higher disorder are better to improve the spin-valley polarization of silicene. This is quite interesting because coincides with the improvement of the spin-valley polarization induced by structural disorder in MSSLs~\cite{PhysRevB.105.115408}.

Finally, Figure~\ref{tatmr} shows the TMR as a function of ${\Delta}_{z}$ for TM- and P-MSSLs, magenta and black curves, respectively. In both systems, there are oscillations whose amplitude grows as the generation increases. Peaks close to ${\Delta}_{z}\approx 2.8 \Delta_{SO}$ have 100\% of TMR. If we focus on the range $2.5{\Delta}_{SO}<{\Delta}_{z}<2.8{\Delta}_{SO}$, we can notice that the peaks corresponding to P-MSSLs are greater than those of the aperiodic case. However, from the third generation onwards the TMR values are better for TM-MSSLs. For instance, for the fifth generation, the TMR around $\Delta_{z}=2.6\Delta_{SO}$ shows an enhancement of 37\% for TM-MSSLs with respect to P-MSSLs.
\begin{figure}
\begin{center}
\includegraphics[width=0.48\textwidth]{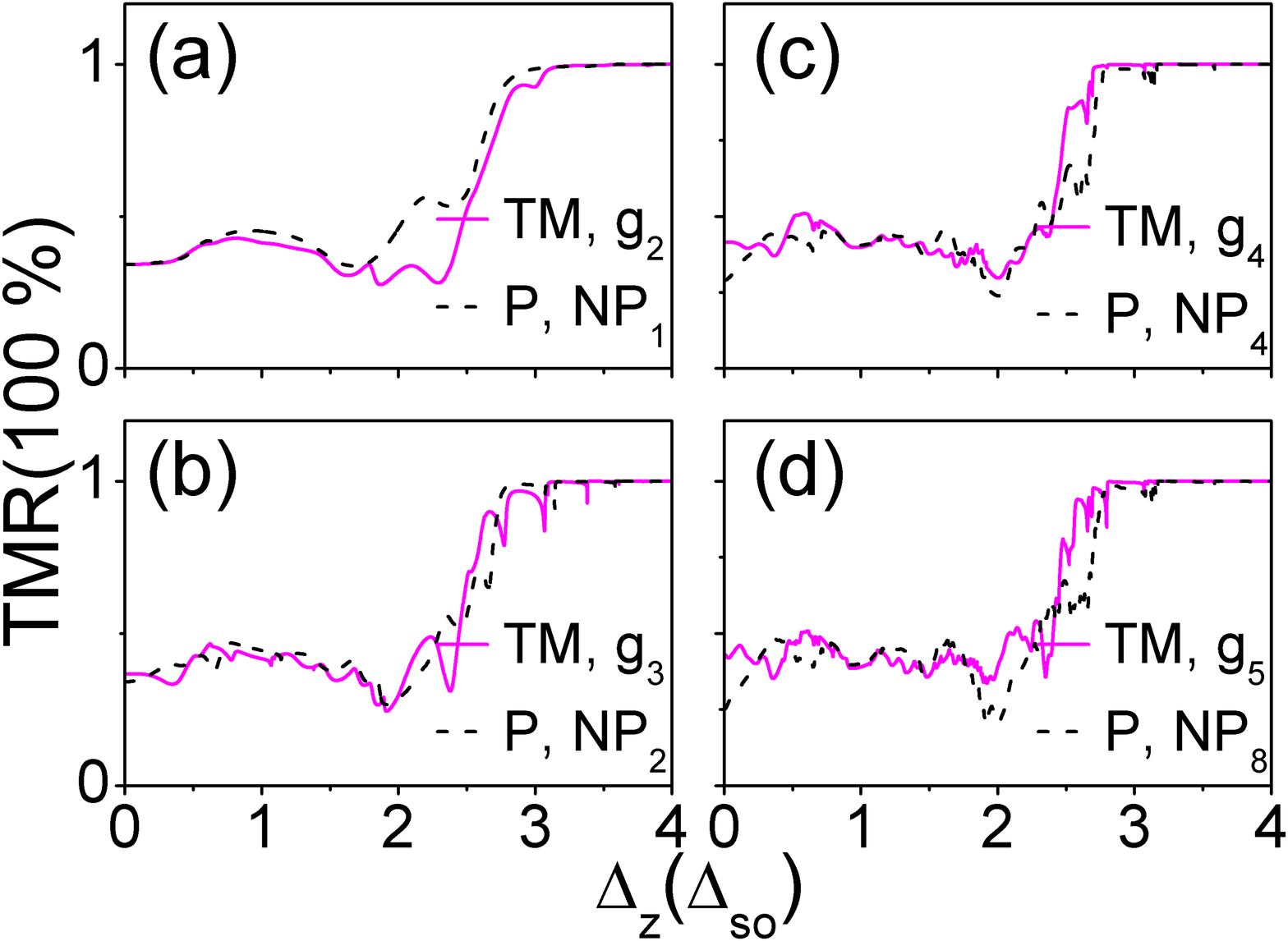}
\caption{\label{tatmr} TMR as function of $\Delta_z$ for (a)-(b) TM- and (c)-(d) P-MSSLs, magenta and black curves, respectively.}
\end{center}
\end{figure}

\section{Conclusions}
In summary, we have studied the spin-valley polarization and TMR in aperiodic MSSLs.  As representative aperiodic superlattices closer to ordered and disordered structures we have assessed F- and TM-MSSLs, respectively.  The charge carriers in aperiodic MSSLs are described as quantum relativistic particles through a low-energy effective Hamiltonian. The transmission and transport properties were obtained within the lines of the transfer matrix method and the Landauer-B\"uttiker formalism, respectively.  We found in general that aperiodicity improves the spin-valley polarization and TMR of MSSLs. In particular,  the spin-valley polarization and TMR are better for TM-MSSLs than for F- and P-MSSLs. The higher disorder associated to TM-MSSLs reduces considerably the conductance oscillations, improving in this way the spin-valley polarization and TMR.  In the case of F-MSSLs the transport properties depend strongly on the parity of the superlattice generation. Superlattices with even generations result in well-defined polarization regions and better TMR values, while superlattices with odd generations are considerably affected by conductance oscillations. In short, aperiodicity could be useful to improve the spin-valleytronic and magnetoresistive properties of MSSLs. 


\begin{acknowledgments}
P.V.-M. acknowledges CONACYT-Mexico for its support in the development of the present work through the scholarship grant No. 658398. I.R.-V. acknowledges CONACYT-Mexico for the financial support through grant A1-S-11655.
\end{acknowledgments}


\end{document}